\documentstyle[aps,prl,twocolumn,psfig]{revtex}
\begin{document}
\preprint{\vbox{\hbox{UCB-PTH-00/19}},
  \vbox{LBNL-46181}}
\newcommand{\rem}[1]{{\bf #1}}
\newcommand{\gev}{{\rm GeV}}
\newcommand{\mev}{{\rm MeV}}
\newcommand{\kev}{{\rm keV}}
\newcommand{\ev}{{\rm eV}}
\newcommand{\cm}{{\rm cm}}
\newcommand{\mpl}{M_{Pl}}
\def\pl#1#2#3{{\it Phys. Lett. }{\bf B#1~}(#2)~#3}
\def\zp#1#2#3{{\it Z. Phys. }{\bf C#1~}(#2)~#3}
\def\prl#1#2#3{{\it Phys. Rev. Lett. }{\bf #1~}(#2)~#3}
\def\rmp#1#2#3{{\it Rev. Mod. Phys. }{\bf #1~}(#2)~#3}
\def\prep#1#2#3{{\it Phys. Rep. }{\bf #1~}(#2)~#3}
\def\pr#1#2#3{{\it Phys. Rev. }{\bf D#1~}(#2)~#3}
\def\np#1#2#3{{\it Nucl. Phys. }{\bf B#1~}(#2)~#3}
\def\d#1{\left[ #1 \right]_D}
\def\f#1{\left[ #1 \right]_F}
\def\a#1{\left[ #1 \right]_A}
\def\VEV#1{\left\langle #1\right\rangle}
\let\vev\VEV
\wideabs{
\title{Neutrino Masses at $v^{3/2}$}
\author{Nima Arkani-Hamed, Lawrence Hall, Hitoshi Murayama, David Smith and Neal Weiner}
\address{
Department of Physics,
University of California,
Berkeley, CA~~94720, USA;\\ 
Theory Group,
Lawrence Berkeley National Laboratory,
Berkeley, CA~~94720, USA}
\date{\today}
\maketitle

\begin{abstract}
Theories in which neutrino masses are generated by a conventional
see-saw mechanism generically yield masses which are $O(v^2)$ in units
where $M_{Pl}=1$, which is naively too small to explain the results
from SuperKamiokande. In supersymmetric
theories with gravity mediated supersymmetry breaking, the fundamental 
small parameter is not $v/M_{Pl}$, but $m_I/M_{Pl}$, where $m_I$ is
the scale of supersymmetry breaking in the hidden sector. We note that 
$m_I^3/M_{Pl}^2$ is only slightly too large to explain
SuperKamiokande, and present two models that achieve neutrino masses
at this order in $m_I$, one of which has an additional suppression
$\lambda_\tau^2$, while the other has additional suppression arising
from a loop factor. The latter model shares a great deal of
phenomenology with a class of models previously explored,
including the possibility of viable sneutrino dark matter. 
\end{abstract}
}
\setcounter{footnote}{0}
\setcounter{page}{1}
\setcounter{section}{0}
\setcounter{subsection}{0}
\setcounter{subsubsection}{0}

\section{Introduction}
Neutrinos are exceedingly light compared to other fermionic
elementary particles.  For instance, the atmospheric neutrino data
\cite{SK} suggest $m_{\nu} 
\simeq 0.06$~eV.  To understand the smallness of neutrino masses, we 
usually invoke the presence of a scale $M$ that is much larger than
$v \approx 200$~GeV, the scale of electroweak symmetry breaking.  This
scale is  
conventionally taken to be roughly the reduced Planck mass,
$M_{Pl} \approx 2 \times 10^{18}\gev$, or the GUT scale, $M_{GUT}
\approx 2 \times 10^{16}\gev$. Setting $M=1$, $v$ becomes 
a small parameter of the theory. Charged fermion masses arise at 
order $v$, while neutrino masses arise at $v^{2}$. Note that any 
particle of mass $\le v^{3}$ is massless for all particle physics 
practicalities, although such masses could be interesting for
cosmology.

The see-saw mechanism\cite{seesaw} 
is certainly the simplest explanation of why 
neutrino masses are so small - however, the result $v^{2}/M$ 
gives neutrino masses which are $\sim 30-3000$ times too light 
to explain the atmospheric neutrino anomaly\cite{SK}, for $M=M_{GUT}$ or
$M_{Pl}$. Frequently this is fixed by
instead taking $M \sim 10^{14}-10^{15}\gev$. 
In this paper we propose an alternative approach.

In supersymmetric theories, a standard framework has supersymmetry 
broken in a hidden sector at the intermediate scale $m_{I} \sim 10^{10}- 
10^{11}\gev$ which allows supergravity mediation of supersymmetry 
breaking to the standard model to generate the weak scale $v \sim 
m_{I}^{2}/M_{Pl}$. We call this the intermediate-scale see-saw. An 
important question is at what order neutrino masses arise within this
framework.  If the relevant operator is $(LH)^2/M_{Pl} $, 
then $m_{\nu} \sim v^{2}/M_{Pl}$ as before. However, with $M=M_{Pl}=1$, 
the small dimensionless parameter of the expansion is now $m_{I}$ 
rather than $v$. There are now more possibilities, namely $m_{\nu}$ 
could occur at order $m_{I}^{3}, m_{I}^{4}, m_{I}^{5}\ldots$ 
corresponding to $v^{3/2}, v^{2}, v^{5/2}\ldots$ We propose that 
scale for atmospheric neutrino oscillations arises at $O(m_{I}^{3})$. 
The scale of solar 
oscillations can also arise at $O(m_{I}^{3})$ with an extra suppression
due to approximate flavor 
symmetries, or it can instead occur at $O(m_{I}^{4})$.  Alternatively,
the hierarchy between the atmospheric and solar scales can be traced to 
an approximately rank one loop integral, as discussed in section 
\ref{sec:radnu}. 

\section{Neutrino masses at $m_{I}^{3}$}
How might neutrino masses arise at $O(m_{I}^{3})$? People have noted 
before that in a framework with large extra dimensions such that 
$M_{Pl}$ is lowered to $m_{I}$, the ordinary see-saw result 
yields $m_{\nu} \sim v^{2}/m_{I}$\cite{edv32}. Such an approach is quite 
distinct from our perspective, where $m_{I}$ is the small parameter.
Other authors
\cite{models} have explored the possibility of generating neutrino
masses from higher dimension operators involving a field with 
a vacuum expectation value (vev) 
at an intermediate scale, such as $10^{11}\gev$. However, in these models,
a supersymmetry-conserving vev generates the
small couplings and masses, and the size of the vev is
essentially a free parameter of
the theory. In this paper, we will instead utilize the direct connection to
supersymmetry breaking explored in \cite{us}.  Adopting this approach 
can have significant phenomenological ramifications, as we will find 
for the model of section \ref{sec:radnu}.

Right handed states, singlet under the standard model, might be light 
if they are protected by some global symmetry $G$, analogous to 
a symmetry 
used to prevent a Planck-scale $\mu H_{u} H_{d}$ term in the 
superpotential\cite{giudicemasiero}. In \cite{us}, it was noted that 
if $G$ is broken in the supersymmetry breaking 
sector, then it is quite natural to have light neutrinos. In the 
models presented in \cite{us}, the neutrino masses arose at $O(m_{I}^{4})$.

We can employ the same framework to generate 
$O(m_{I}^{3})$ masses instead. Consider the superpotential
\begin{equation}
        \f{ X NN + {LNH}}
        \label{eq:simplemodel}
\end{equation}
where $N$ is a standard model singlet and $X$ is a field in the 
supersymmetry breaking sector that takes on an $A$ component vev
$\vev{X}=m_{I}$ (for instance, the superpotential
\begin{equation}
\left[S(X \overline{X}-m_I^2)+{\overline X}^2 Y +X^2{\overline Y}\right]_{F}
\label{eq:or}
\end{equation}
generates $F_S \sim F_Y= F_{\overline Y} \sim m_I^2$ and $A_X =
A_{\overline X} \sim m_I$, but $F_X = F_{\overline X}=0$\cite{chargefootnote}).
The Lagrangian then contains
\begin{equation}
        \left[ m_{I} NN\right]_{F} +
         \left[ LNH \right]_{F},
        \label{eq:modellagr}
\end{equation}
and when the Higgs takes on a vev, we have small Dirac masses for the 
neutrinos in addition to the Majorana mass for the right handed 
neutrino. Setting $M_{Pl}=1$, the neutrino mass matrix is 
\begin{equation}
        m_{LR} = \pmatrix{ 0 & m_{I}^{2} \cr m_{I}^{2} & m_{I}},
        \label{eq:simplematrix}
\end{equation}
leading to a see-saw mass   
$m_{\nu} \approx m_{I}^{3}$ for the left handed neutrino. 
If we further assume that the Dirac masses 
are suppressed by $\lambda_{\tau}$, for instance as might occur with 
flavor symmetries, the neutrino mass is $\lambda_{\tau}^{2} 
m_{I}^{3} \approx 0.1 \ev$.  (A similar estimate $\lambda_\mu^2 m_I^3$
gives an approximately correct mass scale for the LOW solution to the
solar neutrino problem if there are more than one $N$).

This model is exceedingly simple, but it illustrates the fact that 
once we take the small parameter of the theory to be $m_{I}$ rather 
than $v$, there is no {\em a priori} reason why we should expect 
neutrino masses to occur at $m_{I}^{4}$. 

The phenomenology of this model is identical to that of ordinary 
see-saw neutrino physics, except that we do not expect the signals 
that might accompany broken GUT symmetries at the scale 
$10^{14}\gev$. In contrast, the model that we consider next features 
additional weak-scale states and predicts a much richer phenomenology.

\section{Radiative neutrino masses at $m_{I}^{3}$}
\label{sec:radnu}
In \cite{us}, a model was proposed in which the right handed neutrinos 
get masses at $O(m_{I}^2)$, rather than at $O(m_{I})$.  If the 
Yukawa couplings are order one, then all neutrino 
masses are weak scale. As discussed in \cite{us}, 
this is remedied rather simply.

Suppose that a hidden sector field $X$ acquires 
$A$ and $F$ component vevs of $O(m_{I})$ and $O(m_{I}^{2})$, 
respectively. Consider the operators
\begin{equation}
         {1 \over M_{Pl}}\left(\f{ X L N H_{u}} + \d{ X^{\dagger} NN}\right),
        \label{eq:firstpart}
\end{equation}
which yield 
\begin{equation}
        {m_{I}^{2} \over M_{Pl}} [NN]_{F} + {m_{I}\over M_{Pl}} [LNH]_{F} + 
        {m_{I}^{2}\over M_{Pl}}[LNH]_{A}
        \label{eq:firstcomp}
\end{equation}
when $X$ acquires its vevs.  We take there to be just one $N$ superfield.
The first term in (\ref{eq:firstcomp}) generates a weak scale mass 
for $N$, while the second term generates Yukawas of 
order $m_{I}/M_{Pl}$. The neutrino mass matrix which we generate at 
tree level is (again setting $M_{Pl}=1$)
\begin{equation}
        m_{\rm tree} = \pmatrix{0 & m_{I}^{3} \cr m_{I}^{3} & m_{I}^{2}}.
        \label{eq:treemass}
\end{equation}
Although this is not the canonical see-saw, as it involves a Dirac 
mass at $O(v^{3/2})$, it nevertheless yields the light neutrino mass 
of $v^{2}/M_{Pl}$ as usual. The unusual feature is that the right 
handed neutrino is at the weak scale. The operators of
(\ref{eq:firstpart}) are  
easily justified, for instance by 
assuming ordinary R parity (under which $N$ is odd), 
together with an R symmetry where $N$ has R charge $2/3$, $X$ has R 
charge $4/3$ and $L$ and $H_u$ have R charge $0$ \cite{Rcharges}.

\begin{figure}
\vspace{-2 in}        
\centerline{\psfig{file=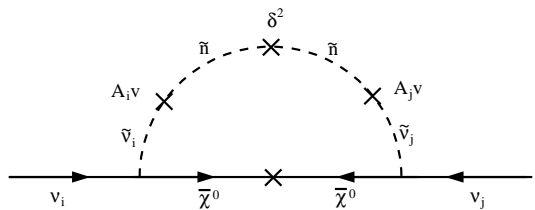,width=.55\textwidth}}
\vspace{-2 in}
\caption{Diagram generating $m_\nu$ at order $m_I^3$.}
\label{fig:loop}
\end{figure}

Once we have included the operator $\d{X^{\dagger} 
NN}$, it is impossible to forbid $\d{ X^{\dagger} X 
X^{\dagger} NN}$\cite{modeldif}. This operator induces 
a lepton number violating scalar mass of the form 
$\delta^{2} \tilde n \tilde n + h.c.$, where $\delta \sim 
m_{I}^{5/2}/M_{Pl}^{3/2}$. The importance of this operator is that it 
allows a radiative contribution to the light neutrino mass at 
$O(m_{I}^{3})$ from the diagram of figure \ref{fig:loop}.

If we take $m_{\tilde \nu_{L}} \approx m_{\tilde n} \approx 
m_{\chi^{0}}$, and call them generically $\tilde{m}$, we generate a
neutrino mass 
\begin{equation}
        m_{\nu} \approx {g^{2} \over 384 \pi^{2}} {A^{2} v^{2} \delta^{2} \over 
        \tilde m^{5}}.
        \label{eq:radmass}
\end{equation}
Taking $A\sim \tilde m \sim v$, this becomes 
\begin{equation}
        m_{\nu} \approx {g^{2} \over 384 \pi^{2}} \left( {m_{I}^{3} \over 
        M_{Pl}^2 } \right).
        \label{eq:radmass2}
\end{equation}
Note that this is {\em larger} than the lighter eigenvalue of 
(\ref{eq:treemass}) 
since it occurs at $O(m_{I}^{3})$ rather than $O(m_{I}^{4})$. A tree 
level neutrino mass at $O(m_{I}^{3})$ is $O(\kev)$ - too heavy to be 
interesting. However, our mechanism automatically leads to a loop 
factor of order $10^{-4}$ giving masses of $O(.1\ev-1 \ev)$ - within one 
order of magnitude of the scale necessary to explain the atmospheric neutrino 
anomaly!

Although the $A$ terms couple $\tilde{n}$ to only a single linear
combination of $\tilde{\nu}$'s, the loop diagram of figure \ref{fig:loop}
can generate more than one neutrino mass eigenvalue.
For incoming $\nu_i$ and $\nu_j$, the value of the loop integral, $L_{ij}$ 
has nontrivial dependence on the corresponding sneutrino masses
$\tilde{m}_i$ 
and $\tilde{m}_j$. 
The resulting mass matrix, $m_{ij} \propto A_i A_j L_{ij}$ is not
necessarily rank one, and we can expect a second and third eigenvalue.
Although $L_{ij}$ does depend on $i$ and $j$, to a large extent it
factors into $f_i f_j$, and is approximately rank one. Consequently, 
the second eigenvalue is
suppressed greatly compared to first. We have investigated this
numerically, and for a broad range of the parameters the second
eigenvalue is typically a factor of $10^{-2}$ or smaller down from the 
first. Consequently, it may either be that the mass scale to explain the
solar neutrino anomaly arises from this additional suppression, 
or that it arises 
at $O(m_I^4)$ from (\ref{eq:treemass}).
Of course, one could alternatively use more than one $N$ and a
hierarchical $A$ matrix to generate a hierarchical $m_{ij}$.

\section{Neutrino Mass Anarchy}

The possibility has been explored elsewhere that neutrino mass
matrices have no ordering structure, such as a flavor
symmetry \cite{anarchy}. 
Absence of flavor symmetry is even {\em more} reasonable in our 
framework - all
suppressions arise naturally via loop factors or factorization, or
occur at different orders in $m_I$. 

If the parameters in the model display no apparent structure, that 
is, the $A_{i}$ are all roughly equal, and likewise the $m_{\tilde 
\nu_{i}}$ are roughly - but not exactly - the same, then we have a 
natural justification for the large mixing observed between 
$\nu_{\mu}$ and $\nu_{\tau}$. We would then expect the solution to the 
solar neutrino problem to similarly involve a large angle: either 
large angle MSW, vacuum oscillations or the LOW solution. 
The only small 
parameter required is $\theta_{13} < 0.16$, required by
CHOOZ\cite{CHOOZ}, but 
we can view this as an accident rather than a fine tuning.

Even if we have a flavor symmetry which explains the structure of the 
charged fermion masses, that would not necessarily preclude such a 
scenario\cite{us}. If the structure of the $A$-terms were determined 
by a supersymmetry-preserving spurion $\lambda_{i}$, i.e.,
\begin{equation}
        A_{i}\tilde n \tilde \nu^{i} h_{u}=
        \frac{\f{X}}{\mpl} \lambda_{i} \tilde n \tilde \nu^{i} h_{u},
\label{eq:susycon}
\end{equation}
then we expect a hierarchy in $A_{i}$ related to that which we find in 
the charged leptons. However, if $X$ carries a flavor index itself, i.e.,
\begin{equation}
        A_{i}\tilde n \tilde \nu^{i} h_{u}=
        \frac{\f{X_{i}}}{\mpl}  \tilde n \tilde \nu^{i} h_{u},
\label{eq:susybre}
\end{equation}
then the situation is quite different. Because the structure of 
$X_{i}$ is determined in the supersymmetry breaking sector, it need 
not be related to the structure of the lepton masses. In this case, 
even with a flavor symmetry, we would expect large angles to arise in 
the neutrino sector.

\section{Phenomenology and cosmology}
The phenomenology of the model presented in section \ref{sec:radnu} 
is very interesting. It is
essentially identical to that explored in \cite{us} for the case of
a single $N$ superfield. The presence of a weak-scale $\tilde{n}$ that
mixes through weak-scale $A$ terms to the left-handed sneutrinos can 
profoundly affect the sneutrino spectrum.  For instance,
a sneutrino mass eigenstate is not subject to the $Z$-width constraint
if it is mostly composed of $\tilde{n}$, and its mass can be different from
that of $\tilde{l}_L$ by far more than just the $D$-term splitting.  The
$A$ terms could potentially induce invisible Higgs decays into light
sneutrino pairs, and $\tilde{\nu} \tilde{l}$ might be the
dominant decay mode for the charged Higgs boson.  Finally, cascade
decays producing heavy sneutrinos that subsequently decay into a Higgs 
and light sneutrino could conceivably be the dominant source of
Higgs production at the LHC. There are numerous other potential
consequences, arising in particular scenarios, which we will not
discuss here. 

The presence of the additional $\tilde{n}$ state revives
the possibility of sneutrino dark matter, 
as was explored in \cite{us} in the context of 
both lepton number conserving and
lepton number violating models. In the lepton number conserving case,
direct detection experiments require $m_{\tilde \nu}< 3$ GeV\cite{dmsearches}. 
In contrast, the lepton number violating scalar mass term central to the
present model allows for easy evasion of this bound.

Direct detection experiments detect ordinary sneutrino dark matter
through $Z$ boson exchange. However, once a lepton number
violating mass term is present, the CP-even state $\tilde \nu_+$ and CP-odd
state $\tilde \nu_-$ are no longer degenerate in mass. Moreover, scalar
couplings to the $Z$ are off-diagonal, i.e., they couple $\tilde \nu_+$ to
$\tilde \nu_-$, but not $\tilde \nu_-$ to $\tilde \nu_-$. The
scattering of $\tilde \nu_-$ off of a nucleus through $Z$ exchange is
kinematically forbidden for $\Delta m > \beta_{h}^2 m_-
m_A/2(m_A+m_-)$, where $\Delta m=m_+-m_-$ is the mass splitting
between the CP-even and CP-odd states, $m_A$ is the mass of the 
nucleus, and $\beta_{h}=10^{-3}$ for virialized halo particles on average.   
Thus, even for sneutrino masses of
$O(100$ GeV) , direct detection limits are essentially harmless,
stipulating only that the lepton number violating mass is adequately
large. For example, taking $m_{\tilde \nu}=100$ GeV and 
$m_A=72$ GeV for a Ge target, we simply need $\Delta m > 20$ keV to prevent
direct detection. Because $\Delta m=\delta^2/m_{\tilde \nu}$, this  
corresponds to $\delta > 45\mev$, which is of the 
order of what we expect from $m_I^{5/2}/\mpl^{3/2}$. 

Sneutrino dark matter will still scatter from the nuclei via Higgs
exchange. The cross section per nucleon for this is small, however, given by
\cite{us}
\begin{equation}
\sigma \approx 1.8 \times 10^{-43} \sin^2 2\theta \hskip1in
\label{eq:dmcs}
\end{equation}
\begin{equation}
\times \left( {A \over
    100\gev} \right)^2 \left({130 \gev \over m_h} \right)^4 \left( {100 
  \gev \over m_{\tilde \nu}} \right)^2 \cm^2,
  \nonumber
\end{equation}
where 
\begin{equation}
\tilde \nu_{DM} = \tilde{\nu} \sin \theta + \tilde{n} \cos \theta.
\label{eq:mix}
\end{equation}
The current upper bound on $\sigma$ for a dark matter candidate with
mass $\sim 100
\gev$ is $10^{-41} \cm^2$\cite{dmsearches}, and this bound is expected to be
lowered by orders of magnitude in the near future\cite{futuredm}.  
Thus, this version of 
sneutrino dark matter may be detectable due to Higgs exchange in 
upcoming direct detection experiments.

Sneutrino dark matter with lepton number violation has been previously 
explored\cite{hmm}, precisely because it can evade direct detection
limits. However a large mass splitting ($\Delta m_{\tilde{\nu}} \sim
\gev$) was necessary to
suppress $\tilde \nu_+ \tilde \nu_-$ coannihilation in the early
universe to yield an appreciable relic density. Here, because the
lightest sneutrino is an admixture of left- and right-handed states,
the overall annihilation rate via MSSM processes is suppressed by an 
additional factor of $\sin^4 \theta$. As discussed in \cite{us}, acceptable
relic abundances are obtained for a broad range of parameters.
For example, for sneutrino masses less than $M_W$, the relic density
is essentially determined by the annihilation rate 
via neutralino exchange, and 
one finds
\begin{equation}
\Omega h^2 \approx \left({M_{\tilde W} \over 100 \gev } \right)^2 \left( 
 {\sin \theta \over 0.16 } \right)^4,
\label{eq:relicdensity}
\end{equation}
where $h$ is the reduced Hubble parameter and $M_{\tilde{W}}$ is the
neutral wino mass. Such a simple approximate formula does not apply
when the sneutrino mass is heavy enough so that production of $W$ and
$Z$ pairs becomes relevant, but the abundance is still promising
for reasonable parameter choices.

\section{Conclusions}
While conventional see-saw models generate neutrino masses
proportional to $v^2$, in theories with gravity mediated supersymmetry 
breaking, it is also possible to generate neutrino masses proportional 
to $v^{3/2}$. Models of this type arise when flavor symmetries protect 
the masses of standard model singlet states, but are 
broken in the supersymmetry breaking sector of the theory.

Additional suppression to these masses can arise from Yukawa-type
suppressions, or from loop factors, resulting in values for the
neutrino mass in accordance with observations from Superkamiokande of
an up/down neutrino asymmetry. 

The model developed in section III, which features a right-handed
sneutrino at the weak scale, 
is phenomenologically rich, with dramatic changes to 
collider signatures and the possibility of sneutrino dark
matter. Although the sneutrinos in this model
evade current detection limits on dark matter, the possibility exists 
for their detection at a future experiment.

\vskip 0.15in
\begin{center}
{\bf Acknowledgements}
\end{center}
\vskip0.1in
We thank R. Rattazzi for pointing out an error in a previous version
of this work.
   This work was supported in part by the U.S.
   Department of Energy under Contracts DE-AC03-76SF00098, in part by   
   the 
   National Science Foundation under grant PHY-95-14797.


\begin{thebibliography}{99}
\bibitem{Rcharges} The other fields in the Supersymmetric Standard
  Model have $R$-charges $Q(0)$, $H_d(0)$, $E(+2)$, $U(+2)$, $D(+2)$.
\bibitem{modeldif} In the original model of \cite{us}, $\d{X^{\dagger} 
NN}$ was not the only means of generating a weak-scale $NN$ mass. 
Alternatively, a gauged $U(1)_{B-L}$, broken at the weak scale,
can give the same mass without introducing significant radiative
corrections to the neutrino mass.
\end{thebibliography}
\end{document}